\documentclass[aip,jcp,amsmath,amssymb]{revtex4-1}
\usepackage[T1]{fontenc}
\usepackage[utf8]{inputenc}
\usepackage{tipa}
\usepackage{lmodern}
\usepackage{dcolumn}
\usepackage[version=3]{mhchem}
\usepackage{graphicx}
\usepackage{paralist}
\usepackage{lineno,xcolor}
\usepackage{cancel}
\usepackage{siunitx}
\usepackage{float}
\usepackage[pdftex = true]{hyperref}

\newcommand{\hjswpbe}{$\omega$PBE}
\newcommand{\hjswpbed}{$\omega$PBE-D3}
\newcommand{\figref}[1]{Fig.~\ref{#1}}
\newcommand{\figsref}[2]{Figs.~\ref{#1} and~\ref{#2}}
\newcommand{\tabref}[1]{Table~\ref{#1}}
\newcommand{\eqnref}[1]{Eq.~\ref{#1}}
\newcommand{\wpbesol}{$\omega$PBEsol}
\newcolumntype{d}[1]{D{.}{.}{#1}}

\newcommand{\rhoa}{\rho_{\alpha}}
\newcommand{\rhob}{\rho_{\beta}}
\newcommand{\taua}{\tau_\alpha}

\newcommand{\rs}{r_\mathrm{s}} 
\newcommand{\rsaa}{r_\mathrm{s}^{\alpha\alpha}} 
\newcommand{\rsab}{r_\mathrm{s}^{\alpha\beta}} 

\newcommand{\hcss}{h_{\mathrm{C}\lambda}^{\sigma\sigma'}}
\newcommand{\hcaa}{h_{\mathrm{C}\lambda}^{\alpha\alpha}}
\newcommand{\hcab}{h_{\mathrm{C}\lambda}^{\alpha\beta}}

\newcommand{\aopp}{a_{\alpha\beta}}
\newcommand{\Aopp}{\mathcal{A}_{\alpha\beta}}
\newcommand{\bopp}{b_{\alpha\beta}}
\newcommand{\Bopp}{\mathcal{B}_{\alpha\beta}}
\newcommand{\copp}{c_{\alpha\beta}}
\newcommand{\dopp}{d_{\alpha\beta}}
\newcommand{\apar}{a_{\alpha\alpha}}
\newcommand{\Apar}{\mathcal{A}_{\alpha\alpha}}
\newcommand{\bpar}{b_{\alpha\alpha}}
\newcommand{\Bpar}{\mathcal{B}_{\alpha\alpha}}
\newcommand{\cpar}{c_{\alpha\alpha}}
\newcommand{\dpar}{d_{\alpha\alpha}}

\newcommand{\ec}{E_\mathrm{C}}
\newcommand{\ecab}{E_\mathrm{C}^{\alpha\beta}}
\newcommand{\ecba}{E_\mathrm{C}^{\beta\alpha}}
\newcommand{\ecaa}{E_\mathrm{C}^{\alpha\alpha}}
\newcommand{\ecbb}{E_\mathrm{C}^{\beta\beta}}

\newcommand{\diff}{\mathrm{d}}

\newcommand{\ra}{\mathbf{r}_1}

\newcommand{\dra}{\diff^3\ra}

\newcommand{\erfc}{\mathrm{erfc}}
\newcommand{\erf}{\mathrm{erf}}

\begin{document}
\title{Range-separated meta-GGA functional designed for noncovalent interactions}


\author{Marcin Modrzejewski}
\email{m.m.modrzejewski@gmail.com}
\affiliation{Faculty of Chemistry, University of Warsaw, 02-093 Warsaw, Pasteura 1, Poland}
\author{Grzegorz Chałasiński}
\affiliation{Faculty of Chemistry, University of Warsaw, 02-093 Warsaw, Pasteura 1, Poland}
\author{Małgorzata M. Szczęśniak}
\affiliation{Department of Chemistry, Oakland University, Rochester,
  Michigan 48309-4477, USA}

\begin{abstract}
The accuracy of applying density functional theory to noncovalent interactions
is hindered by errors arising from low-density regions of interaction-induced
change in the density gradient, error compensation between correlation and exchange functionals, and
dispersion double counting. A new exchange-correlation functional designed
for noncovalent interactions is proposed to address these problems.
The functional consists of the range-separated PBEsol exchange considered in two
variants, pure and hybrid, and the semilocal correlation functional of
Modrzejewski et al. [\textit{J. Chem. Phys.} {\bf 137}, 204121 (2012)]
designed with the constraint satisfaction technique to smoothly connect with a dispersion term. 
Two variants of dispersion correction are appended to the correlation functional:
the atom-atom pairwise additive DFT-D3 model and the density-dependent many-body dispersion
with self-consistent screening (MBD-rsSCS). From these building
blocks a set of four functionals is created to systematically examine the role of
pure versus hybrid exchange and the underlying models for dispersion. The new
functional is extensively tested on benchmark sets with diverse nature and size.
The truly outstanding performance is demonstrated for water clusters of varying size,
ionic hydrogen bonds, and thermochemistry of isodesmic n-alkane fragmentation
reactions. The merits of each component of the new functional are discussed. 
\end{abstract}

\maketitle

\section{Introduction}
DFT is one of few quantum-chemical methods capable of dealing with
problems germane to molecular biology and materials science which
involve electronic structure, yet on a scale too large for ab initio wavefunction
tools. So far, however, the approximate character
of affordable functionals seriously restricts their predictive power in several
important areas, the most prominent ones being related to noncovalently bound systems.
An approximate functional focused on performance for noncovalent interactions is the
subject of this work.

During the past decade, a large effort has been devoted to resolve the deficiencies
 in the description of noncovalent interactions. The progress has been indicated by steady
improvement of statistical errors in databases of noncovalent interactions.\cite{burns2011density,podeszwa2012communication} 
Still, part of this apparent advancement is a result of error cancellation between the
dispersion-free part of a functional and its a posteriori dispersion correction. 
Consequently, even for the best performing methods, there exist systems for which the
cancellation does not occur and error spikes beyond the average levels. Examples
of such problematic systems are water clusters studied in this work.

A practical chemist copes with the issue of large, unpredictable errors by cross-checking her
 calculations with several independent approximate functionals. Thus, to make DFT a dependable tool,
we still need new functionals developed independently from the currently existing ones
 and built from well-defined components, which do not exploit obscure error cancellation.

This work introduces a set of new DFT exchange-correlation functionals intended primarily for noncovalent interactions.
They are composed of the recent meta-GGA correlation developed by \citet{modrzejewski2012first},
the range-separated PBEsol exchange\cite{perdew2008restoring,csonka2008improved,henderson2008generalized}
({\wpbesol}), and a dispersion correction, 
 \begin{equation}
 E_\mathrm{XC} = E_\mathrm{C} + E_\mathrm{X}(\omega\text{PBEsol}) + E_\mathrm{disp}. \label{eqn-functional-def}
 \end{equation}

The two variants of the dispersion correction employed in this work are DFT-D3 by \citet{grimme2010consistent} 
(abbreviated as D3) and MBD-rsSCS by \citet{ambrosetti2014long} (abbreviated as MBD). There are
other possible ways of including dispersion not explored
 here.\cite{lee2010higher,vydrov2010nonlocal,sato2010local,becke2005exchange} Furthermore, we assess two
 variants of short-range exchange: pure PBEsol and a hybrid with an addition
 of the short-range HF exchange. For brevity, the full exchange will be called
either \emph{pure} or \emph{hybrid} depending on the fraction of the short-range exact exchange. In total, there are
four combinations of the exchange and dispersion components: MCS-D3, MCS-MBD, MCSh-D3, and MCSh-MBD,
where the first part of the label denotes the exchange approximation (MCS for the pure exchange
and MCSh for the hybrid) and the second part specifies the dispersion correction.
This set of functionals will be collectively referred to as MCS.

The MCS functionals are designed to overcome several issues of the currently available
exchange-correlation approximations.

First, a part of the difficulties in the description of noncovalent systems can be pinpointed to
the poor behavior of approximate exchange functionals in the low-density regions where the density gradient
changes substantially upon bond formation.\cite{lacks1993pair,murray2009investigation} The emergence
of such regions is the signature of noncovalent bonding\cite{johnson2010revealing} and is the source
of major contributions to the interaction energy.\cite{lacks1993pair,murray2009investigation}
For example, depending on the limit of an exchange enhancement factor for large reduced gradients,
the exchange-only interaction curve of a noble gas dimer can be either attractive (as in PBE) or
much more repulsive than the Hartree-Fock limit (as in B88).\cite{kannemann2009van} Although the behavior
of the exchange is not decisive for the performance of the full exchange-correlation functional due to
the possible error cancellation, it may obscure the interpretation of interaction energies and eventually worsen
the compatibility with dispersion corrections. One way of ensuring that DFT exchange-only interaction curves
resemble the Hartree-Fock ones is by employing range-separated exchange functionals.\cite{kamiya2002density}
Also inclusion of the exact second-order gradient expansion of the exchange functional improves description
of the regions relevant for noncovalent systems.\cite{csonka2008improved,johnson2012density} 
The $\omega$PBEsol exchange included in the MCS functionals combines both of these remedies.

The second problem with the existing DFT treatments of noncovalent interactions
is that a dispersion correction, such as D3, tends to disguise the shortcomings of the base semilocal
functional. This may lead to an inconsistency that the \emph{long-range} dispersion correction calibrated for
an underbinding semilocal functional becomes larger than the reference value of the \emph{total} dispersion as
obtained from the SAPT approach.\cite{misquitta2005intermolecular}
A BLYP-D3 treatment of complexes from the S22 database serves as an example of such an inconsistency. We discuss
this issue later in the text.

The third possible source of errors is double counting of short-range correlation
by a semilocal correlation functional and a dispersion correction. The semilocal correlation
model employed here is designed to avoid this issue via the design of the corresponding
correlation hole. The hole is equipped with a single empirical parameter to control
its range. To eliminate the overlap with the dispersion correction,
the damping of the hole for large $r_{12}$ is adjusted through empirical optimization.\cite{modrzejewski2012first}

Some of the features of a density functional deemed here important for noncovalent systems
have been recognized and built into the $\omega$B97X-D\cite{chai2008long} and
$\omega$B97X-D3\cite{lin2013long} functionals. Both of these models employ range-separated
exchange and have $15$ empirical parameters in their energy expressions optimized simultaneously
with the dispersion corrections. (A systematic analysis
of the B97-type functionals has demonstrated, however, that the number of empirical parameters
should be reduced to improve the performance outside the training sets.\cite{mardirossian2014exploring})
The dispersionless density functional of \citet{pernal2009dispersionless} is also an example 
of a heavily parametrized functional designed to be used in combination with a dispersion term.

\section{Theory}
\subsection{Semilocal correlation}
The first term of \eqnref{eqn-functional-def}, $E_\mathrm{C}$, stands for the recently
proposed correlation functional of \citet{modrzejewski2012first} The functional has been derived starting
from a meta-GGA model for the spin-resolved correlation hole,
\begin{align}
  \hcab(\ra,r_{12}) &= (\aopp + \bopp r_{12} + \copp r_{12}^2) \exp(-\dopp r_{12}), \label{opphole-def}\\
  \hcaa(\ra,r_{12}) &= r_{12}^2 (\apar + \bpar r_{12} + \cpar r_{12}^2) \exp(-\dpar r_{12}), \label{parhole-def}
\end{align}
where $a_{\sigma\sigma'}$, $b_{\sigma\sigma'}$, and $c_{\sigma\sigma'}$ are functions of density
at a given point, obtained from analytic formulas for the short-range (small $r_{12}$) part of the pair 
correlation function in the homogeneous electron gas.\cite{gori2001short,rassolov2000reply} 
These formulas were modified to include dependence on the kinetic energy density
to eliminate the spurious self-interaction in the parallel-spin part.\cite{modrzejewski2012first}
The only empirical parameter of the correlation model, $G$, governs the exponential damping,
\begin{align}
  \dopp &= \frac{2.1070}{\rs^{\alpha\beta}} + d_\mathrm{grad}, \label{dopp-def} \\
  \dpar &= \frac{2.6422}{\rs^{\alpha\alpha}} + d_\mathrm{grad}, \\
  d_\mathrm{grad} &= \frac{G}{\rs} \frac{\nabla\rho \cdot \nabla\rho}{\rho^{8/3}}. \label{dgrad-def}
\end{align}
The larger the numerical value of $G$, the more short-ranged character
of $\hcss(\ra,r_{12})$. Thus, $G$ can be optimized to adjust the range of the approximate
correlation hole to complement, in the manner that avoids double counting, the long-range correlation
contributed by the selected variant of a dispersion correction. It should be emphasized that all the exact constraints 
that are built into our correlation model are obeyed when varying the value of $G$.\cite{modrzejewski2012first}
In particular, the short-range Taylor expansion of $\hcss(\ra,r_{12})$, which has been accurately modeled after
the homogeneous electron gas,\cite{gori2001short} remains unchanged when tuning the correlation functional to a specific
dispersion correction and an exchange functional. Put differently, the empirical adjustment applied to merge
the long-range dispersion with the semilocal correlation does not adversely affect the features which are
reliable already at the semilocal level.\cite{burke1998semilocal}

For reader's convenience, we present $E_\mathrm{C}$ in a form ready for implementation. Following Ref.~\citenum{modrzejewski2012first},
 $E_\mathrm{C}$ is represented as a sum of spin-parallel
 and antiparallel components:
 \begin{equation}
   \ec = \ecab + \ecba + \ecaa + \ecbb.
 \end{equation}
As for any semilocal functional, $E_\mathrm{C}$ is evaluated by numerically integrating
the density of the correlation energy on a molecular grid,
\begin{align}
   \ecab &= \frac 1 2 \int \dra \int_0^1 \diff \lambda \int_0^\infty  \frac{\rho_\alpha(\ra)\hcab(\ra,r_{12})}{r_{12}} 4\pi r_{12}^2 \diff r_{12} \nonumber \\
         &= \int \diff^3 \ra \rhoa \pi \frac{\Bopp + \Aopp\dopp}{\dopp^3}, \label{ecab-def} \\ 
   \ecaa &= \frac 1 2 \int \dra \int_0^1 \diff \lambda \int_0^\infty \frac{\rho_\alpha(\ra)\hcaa(\ra,r_{12})}{r_{12}} 4\pi r_{12}^2 \diff r_{12} \nonumber \\
         & = \int \diff^3 \ra \rhoa \pi \frac{8\Bpar +  4\Apar\dpar}{\dpar^5}.
\end{align}
The integral over the coupling constant $\lambda$ is done analytically.
$\Aopp$, $\Bopp$, $\Apar$, $\Bpar$, $\dopp$, and $\dpar$ are functions
 evaluated at each grid point,
\begin{align}
   \Aopp &= \frac{\rhob}{\rsab} \left[ \left(-P_0 + \sum_{k=1}^4 P_k
     (\rsab)^k \right)\exp\left( -P_5 \rsab \right) + P_0 \right] -
   \rhob, \label{Aopp-def}\\
   \Bopp &= \frac{\rhob}{(\rsab)^2} \left[ \left(-Q_0 + \sum_{k=1}^5 Q_k (\rsab)^k
     \right) \exp\left(-Q_6 \rsab \right) + Q_0 \right] + \dopp \Aopp,
   \label{Bopp-def}\\
   \Apar &= \frac{D_\alpha}{3 \rsaa}\left[ \left( -R_0 +
     \sum_{k=1}^2 R_k (\rsaa)^k \right) \exp\left( -R_3 \rsaa \right)
     + R_0 \right] - \frac{D_\alpha}{3}, \label{Apar-def}\\
   \Bpar &= \frac{D_\alpha}{6(\rsaa)^2} \left[ \left( -S_0 +
     \sum_{k=1}^3 S_k (\rsaa)^k \right)\exp\left( -S_4 \rsaa \right) +
     S_0 \right] + \dpar \Apar \label{Bpar-def}.
 \end{align}
The symbols in Eqs.~\ref{dopp-def}-\ref{dgrad-def} and Eqs.~\ref{ecab-def}-\ref{Bpar-def}
are defined as follows: $\rhoa$ and $\rhob$ are
 electronic spin-densities; $\rho$ is the total electronic density; $\taua$ is the kinetic energy density,
\begin{equation}
\taua = \sum_i^{N_\alpha} |\nabla\psi_{i\alpha}|^2;
\end{equation}
the variable $D_\alpha$ appearing in the parallel-spin part depends on the electron density, its gradient, and $\taua$,
\begin{equation}
D_\alpha = \taua - \frac{|\nabla\rho_\alpha|^2}{4\rhoa}; \label{d-inhom}
\end{equation}
$\rsaa$, $\rsab$, and $\rs$ depend only on electron (spin)-densities:
\begin{align}
  \rsaa &= \frac{\left( 3/\pi \right)^{1/3}}{2
    \rhoa^{1/3}}, \label{rsaa-def} \\
  \rsab &= \frac{\left( 3/\pi \right)^{1/3}}{\rhoa^{1/3} +
    \rhob^{1/3}},  \label{rsab-def} \\
  \rs &= \left( \frac{3}{4\pi\rho} \right)^{1/3}.
\end{align}
$\ecbb$ is obtained by relabeling the spin indices in $\ecaa$; note also that the equality
$\ecab=\ecba$ holds.  The only empirical parameter in the correlation functional is
$G$ (see \tabref{empirical-constants}). The nonempirical parameters appearing in Eqs.~\ref{Aopp-def}-\ref{Bpar-def}
are derived from a short-range model of the correlation hole in the
homogeneous electron gas.\cite{modrzejewski2012first} Their values are defined
in \tabref{abinitio-constants}.

\begin{table}
  \caption{Ab initio numerical constants appearing in Eqs.~\ref{Aopp-def}--\ref{Bpar-def}.}
  \label{abinitio-constants}
  \begin{tabular}{l|D{.}{.}{10} D{.}{.}{10} D{.}{.}{10} D{.}{.}{10}}
    \hline \hline
    k &     P_k    &   Q_k      & R_k       &   S_k \\
    \hline
    $0$ &  1.696                 &  3.356               &  1.775                &  3.205              \\
    $1$ & -0.2763                & -2.525               &  0.01213              & -1.784              \\
    $2$ & -0.09359               & -0.4500              & -4.743\times 10^{-3}   &  3.613\times 10^{-3} \\
    $3$ &  3.837\times 10^{-3}    & -0.1060              &  0.5566               & -4.743\times 10^{-3} \\
    $4$ & -2.471\times 10^{-3}    &  5.532\times 10^{-4}  &                       &  0.5566             \\
    $5$ &  0.7524                & -2.471\times 10^{-3}  &                       &                     \\
    $6$ &                        &  0.7524              &                       &                     \\
    \hline \hline
  \end{tabular}
\end{table}

\begin{table}
 \caption{Empirical parameters of the four tested MCS functionals. The columns
labeled ``MCS'' and ``MCSh'' correspond to the pure and hybrid variants
of the exchange, respectively.} \label{empirical-constants}
 \begin{tabular}{llll}
\hline\hline
name         & MCS          & MCSh        & definition \\
\hline
$G$          & 0.075        & 0.100       & \eqnref{dgrad-def} \\
$\omega$     & 0.300        & 0.200       & \eqnref{eqn-rangeseparation}\\
$\alpha$     & 0.000        & 0.200       & \eqnref{alpha-def} \\
\hline
\multicolumn{4}{l}{D3 dispersion} \\
$r_6$        & 1.1822       & 1.2900  & \eqnref{grimme-damp} \\
$s_8$        & 0.7740       & 1.3996  & \eqnref{grimme-energy} \\
\hline
\multicolumn{4}{l}{MBD dispersion} \\
$\beta$      & 0.8033       & 0.7242  & \eqnref{mbd-damping} \\
\hline\hline
 \end{tabular}
\end{table}

\subsection{Dispersion correction}
The semilocal exchange-correlation functional is supplemented with a dispersion
correction. To confirm the versatility of our approach, we assess
two models of the dispersion interaction: D3\cite{grimme2010consistent} and MBD.\cite{ambrosetti2014long}

The dispersion energy in the D3 approximation is defined as
\begin{align}
 E_\mathrm{disp}(\text{D3}) &= \sum_{AB} \sum_{n=6,8} s_n \frac{C^{AB}_n}{R^n_{AB}} f_{n}\left(R_{AB}\right), \label{grimme-energy} \\
 f_n\left(R_{AB}\right) &= \frac{1}{1+6(R_{AB} / (r_n R^{AB}_0))^{-\alpha_n}}. \label{grimme-damp}
\end{align}
The D3 model contains two empirical, functional-dependent
parameters:\cite{grimme2010consistent} $s_8$ and $r_6$. Other parameters appearing in Eqs.~\ref{grimme-energy} and~\ref{grimme-damp}
are common to all functionals and are defined in Ref.~\citenum{grimme2010consistent}. The $C_6^{AB}$ coefficients are interpolated
from the ab initio tabulated data obtained for hydrides.\cite{grimme2010consistent} The main advantage of D3 is that it is
thoroughly tested\cite{burns2011density} and available in almost any quantum-chemical program.
 It also offers simple to compute derivatives with respect to nuclear coordinates,
which is important for structure optimizations.

The second considered model of the dispersion interaction is MBD.\cite{ambrosetti2014long}
A computation of the MBD energy requires two steps. First, the screening equation\cite{tkatchenko2012accurate}
is solved for frequency-dependent polarizabilities. Then, the solution of the screening
 equation is used to set up the Hamiltonian of interacting quantum
harmonic oscillators whose correlation energy models the long-range dispersion energy of
the real system. While the computational cost of MBD is larger than that of D3, it is still
negligible compared to the SCF step. The dipole interaction in the screening equation as well as in the
MBD Hamiltonian is range separated with a Fermi-type damping function:\cite{ambrosetti2014long}
\begin{align}
f_\mathrm{MBD}(R_{AB}) &= \frac{1}{1 + \exp\left[ - 6 \left( R_{AB} / S_\mathrm{vdW} - 1 \right) \right]}, \\
S_\mathrm{vdW} &= \beta (R_\mathrm{vdW}^A + R_\mathrm{vdW}^B), \label{mbd-damping}
\end{align}
where $\beta$ is the only functional-dependent parameter of the model. The MBD model is expected
to be a good approximation for large molecular systems where an atom-pairwise approximation may no
longer capture the many-body contributions to the total interaction energy.\cite{ruzsinszky2012van,gobre2013scaling}

\subsection{Exchange}
The exchange functional is composed of the short-range
{\wpbesol} exchange,\cite{henderson2008generalized} long-range HF exchange, and optionally
a fraction $\alpha$ of the short-range HF exchange,
\begin{equation}
E_\mathrm{X,SR} = (1-\alpha) E_\mathrm{X,SR}(\omega \mathrm{PBEsol}) + \alpha E_\mathrm{X,SR}(\mathrm{HF}). \label{alpha-def}
\end{equation}
The short-range and long-range parts of $E_\mathrm{X}$ are defined through
the decomposition of the $1/r_{12}$ operator,
\begin{equation}
 \frac{1}{r_{12}} = \frac{\erfc\left(\omega r_{12}\right)}{r_{12}} + \frac{\erf\left(\omega r_{12}\right)}{r_{12}}.
 \label{eqn-rangeseparation}
\end{equation}
We test two variants of the exchange functional: pure ($\alpha=0$ and $\omega=0.3$)
and hybrid ($\alpha=0.2$ and $\omega=0.2$). The range separation parameter $\omega$ for the pure
variant is obtained via empirical optimization. The parameters $\alpha$ and $\omega$
of the hybrid variant are assumed the same as for the LRC-$\omega$PBEh
functional of \citet{rohrdanz2009long} It is worthwhile to note that whereas the fixed value of $\omega$
is convenient in practical computations, the optimal $\omega$ depends on the system size and electronic structure,
which is especially important for donor-acceptor systems.\cite{modrzejewski2013density,koppen2014density}

The PBEsol exchange,\cite{perdew2008restoring} which is the basis for {\wpbesol}, 
has the exact second-order gradient expansion. This feature is important
for solids\cite{perdew2008restoring} and for 
large organic molecules.\cite{csonka2008improved} In contrast to PBEsol, the gradient expansion of the PBE
exchange is not exact; it is designed to cancel the gradient term of the PBE correlation,\cite{csonka2008improved}
which makes it less suitable than PBEsol in conjunction with our correlation functional.

\section{Technical details}
The functionals employed in this study, besides the MCS functionals, 
 are {\hjswpbe},\cite{perdew1996generalized,henderson2008generalized}
 B3LYP,\cite{stephens1994initio} BLYP,\cite{gill1992performance} M06,\cite{zhao2008m06}
 M06-2X,\cite{zhao2008m06} and $\omega$B97X-D.\cite{chai2008long}
$\omega$PBE has its range separation parameter fixed at $\omega=0.4$. The suffix ``-D3'' denotes
functionals with added Grimme's D3 correction.\cite{grimme2010consistent} For water 16-mers and for the S22
database the energies are obtained with the LC-$\omega$PBE functional\cite{vydrov2006importance}
instead of {\hjswpbe}. We supply {\hjswpbe} and LC-$\omega$PBE with the same D3 correction
calibrated by \citet{grimme2010consistent} All DFT computations
 employ the def2-TZVPPD basis\cite{weigend2005balanced,schuchardt2007basis} unless noted otherwise.
The acronyms used to name the types of errors are: mean-absolute percentage deviation (MAPD),
root-mean-square deviation (RMSD), mean absolute deviation (MAD), and mean signed deviation (MSD).
Energies are given in kcal/mol.

The training set for the MCS-D3 and MCSh-D3 functionals is composed of
the noncovalent interactions database of Zhao and Truhlar.\cite{zhao2005design,zhao2005benchmark}
The training database is partitioned into subsets according to the nature of the represented
interactions. The subsets are as follows: WI7/05 (small, dispersion-dominated complexes), 
PPS5/05 ($\pi$-electron dispersion interactions),
DI6-04 (dispersion and dipole interactions), HB6/04 (hydrogen bonds), and CT7/04 (ground-state
charge-transfer interactions).

The optimization of the MCS-D3 functional consisted of the following steps.
First, we generated a grid of parameters $(\omega, G)$ satisfying
$0.100 \le \omega \le 0.450$ and $0.050 \le G \le 0.150$; for each pair $(\omega, G)$, we optimized
 the D3 correction by finding the pair $(r_6, s_8)$ which minimized the objective function
\begin{equation}
 F(r_6, s_8; \omega, G) = 10 \times \text{RMSD(WI7/05)} + \text{RMSD(other)},
\end{equation}
where RMSD(X) denotes the root-mean-square deviation within the subset X of the training set.
Finally, we selected the parameters $(\omega, G, r_6, s_8)$ corresponding
the smallest MAD and MAPD. \tabref{nib-energies} presents the results for the training set.

For the MCS-MBD functional we kept the same values of $\omega$ and $G$ as for MCS-D3.
The only difference is that the parameter $\beta$ of the MBD dispersion was obtained
by minimization of RMSD for the S22 set.\cite{jurecka2006benchmark} The different choice 
of the training sets for the D3 and MBD corrections was due to the poor behavior of the latter for 
small dispersion-bound dimers.

For the hybrid MCS functionals, MCSh-D3 and MCSh-MBD, we did not optimize $\alpha$ and $\omega$,
but fixed these parameters at the same values as in the LRC-$\omega$PBEh functional.\cite{rohrdanz2009long}
The parameter $G$ in the semilocal correlation and the dispersion corrections
were optimized in the same way as for MCS-D3 and MCS-MBD.

\begin{table}
\caption{Training database for the MCS-D3 and MCSh-D3 functionals.
The interaction energies are grouped into five subsets:
WI7/05, DI6/04, CT7/04, HB6/04, and PPS5/05.
The reference CCSD(T) energies are taken from Ref.~\citenum{pernal2009dispersionless}.
The monomer geometries are held rigid at their dimer values. The units are kcal/mol.} \label{nib-energies}
\resizebox{0.4\textwidth}{!}{
\begin{tabular}{llll}
\hline\hline
Dimer                    & CCSD(T) & MCS-D3 & MCSh-D3 \\
\hline                               
\ce{He \bond{...} Ne}    &-0.041   & -0.020 & -0.033 \\
\ce{He \bond{...} Ar}    &-0.058   & -0.016 & -0.034 \\
\ce{Ne \bond{...} Ne}    &-0.086   & -0.015 & -0.048 \\
\ce{Ne \bond{...} Ar}    &-0.131   & -0.039 & -0.061 \\
\ce{CH4 \bond{...} Ne}   &-0.18    & -0.18  & -0.18  \\
\ce{C6H6 \bond{...} Ne}  &-0.41    & -0.41  & -0.52  \\
\ce{CH4 \bond{...} CH4}  &-0.53    & -0.55  & -0.47  \\
\hline
\ce{H2S \bond{...} H2S}  &-1.62    & -1.40  & -1.52 \\
\ce{HCl \bond{...} HCl}  &-1.91    & -1.54  & -1.68 \\
\ce{HCl \bond{...} H2S}  &-3.26    & -3.18  & -3.32 \\
\ce{CH3Cl \bond{...} HCl}&-3.39    & -3.08  & -3.26 \\
\ce{CH3SH \bond{...} HCN}&-3.58    & -3.54  & -3.70 \\
\ce{CH3SH \bond{...} HCl}&-4.74    & -4.94  & -5.13 \\
\hline
\ce{C2H4 \bond{...} F2}  &-1.06    & -0.98  & -1.06 \\
\ce{NH3 \bond{...} F2}   &-1.80    & -1.93  & -1.95 \\
\ce{C2H2 \bond{...} ClF} &-3.79    & -3.74  & -3.93 \\
\ce{HCN \bond{...} ClF}  &-4.80    & -4.15  & -4.03 \\
\ce{NH3 \bond{...} Cl2}  &-4.85    & -4.86  & -5.07 \\
\ce{H2O \bond{...} ClF}  &-5.20    & -5.12  & -5.10 \\
\ce{NH3 \bond{...} ClF}  &-11.17   & -13.65 & -13.89 \\
\hline
\ce{NH3 \bond{...} NH3}  &-3.09    & -2.78  & -2.77 \\
\ce{HF \bond{...} HF}    &-4.49    & -4.06  & -4.13 \\
\ce{H2O \bond{...} H2O}  &-4.91    & -4.60  & -4.61 \\
\ce{NH3 \bond{...} H2O}  &-6.38    & -6.35  & -6.29 \\
\ce{(HCONH2)2}           &-15.41   & -15.02 & -15.14 \\
\ce{(HCOOH)2}            &-17.60   & -18.10 & -17.94 \\
\hline
\ce{(C2H2)2}             &-1.36    & -1.24  & -1.29 \\
\ce{(C2H4)2}             &-1.44    & -1.59  & -1.54 \\
sandwich \ce{(C6H6)2}    &-1.65    & -1.58  & -1.51 \\
T-shaped \ce{(C6H6)2}    &-2.63    & -2.79  & -2.86 \\
displaced \ce{(C6H6)2}   &-2.59    & -2.80  & -2.49 \\
\hline\hline
\end{tabular}
}
\end{table}

To determine the stabilization energy upon complex formation, two definitions
 are employed: the interaction energy and the binding energy.
The interaction energy is defined as
\begin{equation}
E_\mathrm{int} = E\left( \text{dimer AB} \right) - E\left( \text{isolated A} \right) - E\left( \text{isolated B} \right),
\end{equation}
where the monomer geometries are held rigid at their dimer values, and the counterpoise correction is
employed. In the case of water clusters, we use the binding energy instead of $E_\mathrm{int}$,
 \begin{equation}
   E_\mathrm{bind} = E\left( \ce{(H2O)_n} \right) - n E\left( \ce{(H2O)_{isolated}} \right),
 \end{equation}
where the coordinates of water molecules relax upon dissociation from the cluster.
$E_\mathrm{bind}$ does not include the energy of zero-point vibrations. The basis set employed
 for the isolated \ce{H2O} monomers does not include any functions centered on the ghost centers.

The reference binding energies of water 16-mers were obtained by combining $\Delta E_\mathrm{CCSD(T)}$ with the extrapolated
 binding energies at the MP2 level, as proposed by \citet{rezac2011s66},
\begin{equation}
E_\mathrm{bind,CCSD(T)}^{CBS(AVTZ\rightarrow AVQZ)} = E_\mathrm{HF}^{AVQZ} + E_\mathrm{MP2}^{CBS(AVTZ\rightarrow AVQZ)} + \underbrace{\left( E_\mathrm{CCSD(T)}^{AVTZ} - E_\mathrm{MP2}^{AVTZ} \right)}_{\Delta E_\mathrm{CCSD(T)}}, \label{ccsdt-extrap}
\end{equation}
where AVTZ and AVQZ stand for the aug-cc-pVTZ and aug-cc-pVQZ bases.\cite{schuchardt2007basis}
We employed the extrapolation scheme of \citet{halkier1999basis},
\begin{equation}
 E^{CBS(X\rightarrow X+1)}_\mathrm{MP2} = \frac{(X+1)^3 E_\mathrm{MP2}^{X+1} - X^3 E_\mathrm{MP2}^X}{(X+1)^3 - X^3} \label{mp2-extrap}
\end{equation}
with the aug-cc-pVTZ and aug-cc-pVQZ basis sets ($X=3$).
$E_\mathrm{MP2}^{CBS(AVTZ\rightarrow AVQZ)}$ was computed with NWChem\cite{valiev2010nwchem}
 within the resolution-of-identity approximation (RI-MP2) and with the oxygen $1s$ orbitals frozen.
 $E_\mathrm{CCSD(T)}^{AVTZ}$ and $E_\mathrm{MP2}^{AVTZ}$
were taken from \citet{yoo2010high} These contributions do not employ the RI approximation.

\section{Numerical results and discussion}
This section is split into four parts covering a broad spectrum of possible applications.
 \begin{inparaenum}[(i)] \item We begin with two databases of noncovalent
 systems (S22 and A24) which are typical tests for methods focused on
noncovalent interactions.\cite{burns2011density,li2014quantum} \item Next, we turn to water
 clusters of increasing size to test how the accuracy of our method changes when going from small dimers
to clusters with a large number of distant-neighbor interactions. \item We assess the performance of the MCS functionals
for ionic hydrogen-bonded interactions, which is a common motif in biological systems. \item Finally, we
focus on the isodesmic reaction of n-alkanes, which is a well-known case where approximate functionals
fail to fully account for the effect of intramolecular noncovalent interactions.
\end{inparaenum}

\subsection{S22 and A24 databases}
\begin{figure}
\includegraphics[width=1.0\textwidth]{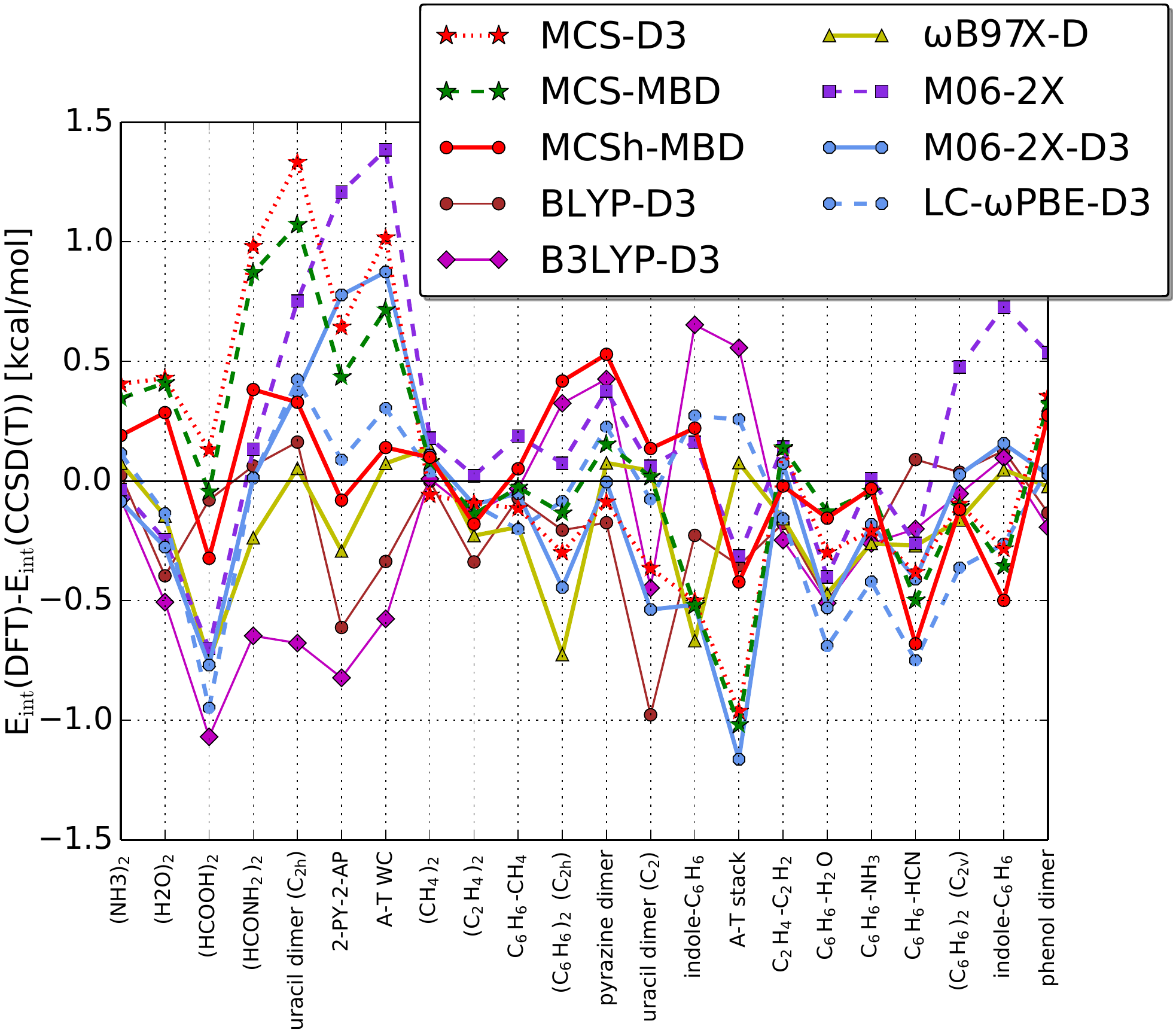}
\caption{Detailed results for the S22 database. The deviations are with
 respect to the CCSD(T) results of \citet{podeszwa2010improved} The energies for the functionals
other than MCS are taken from \citet{goerigk2011thorough}. 2-PY-2-AP
denotes 2-pyridone\ldots 2-aminopyridine.} \label{fig-s22}
\end{figure}

S22 and A24 are two databases of noncovalent dimers which facilitate 
comparisons of density-functional approximations.\cite{jurecka2006benchmark,rezac2013describing} The molecules
contained in these databases are listed in \figsref{fig-s22}{fig-a24}. We compare the MCS functionals against the
leading functionals in the field of noncovalent interactions:\cite{burns2011density,li2014quantum}
Minnesota-family functionals M06-2X and M06-2X-D3, dispersion-corrected range-separated hybrid $\omega$B97X-D,
and two functionals based on the B88 exchange:\cite{becke1988density} B3LYP-D3 and BLYP-D3.

An inspection of \tabref{tab-s22a24} shows that all the MCS functionals afford
small percentage errors within the S22 database. Notably, the two hybrids, MCSh-D3 and MCSh-MBD, have errors
below $6\%$. The pure variants, MCS-D3 and MCS-MBD, tend to underbind the dimers from the hydrogen-bonded
subset of S22 (see \figref{fig-s22}). For the formamide
and uracil dimers the error is the most pronounced and reaches about \SI{1}{kcal\per mol}. The underbinding is
eliminated completely only when both the hybrid exchange and MBD correction are employed
simultaneously. The resulting functional, MCSh-MBD, has exceptionally small absolute as well as relative errors.

Contrary to the results for the S22 database, for A24 we observe that substituting MBD for D3
worsens the percentage errors. This is observed especially for small systems weakly bound by dispersion: \ce{(CH4)2},
\ce{Ar\bond{...}CH4}, and \ce{Ar\bond{...}C2H4}. We stress, however, that these are the only cases where
MBD is systematically inferior to D3.

Although the functionals based on the B88 exchange,\cite{becke1988density} B3LYP-D3 and BLYP-D3,
yield excellent total interaction energies for the S22 database, the physical content of these energies is troubling.
It has been known since the work of \citet{lacks1993pair} that B88 is a
much more repulsive exchange component than the exact HF exchange.
To cancel this contribution, a massive attractive term must be added to the interaction energy. Indeed,
the D3 correction for the B88-based functionals tends to be tens of percent larger
than $E_\mathrm{disp}$ for the MCS-D3 functional (see \tabref{tab-disp}).

While it is impossible to ascertain the precise, physically-sound amount of the D3 correction,
we argue that a large part of the dispersion contribution for the B88-based functionals serves only to cancel the 
overrepulsive exchange. D3 is based on the asymptotic multipole form of the
dispersion term defined in SAPT (\eqnref{grimme-energy}). Thus, it accounts only for
 the \emph{long-range} part of the dispersion interaction, and cannot, for the equilibrium dimers of S22,
 be as large as the \emph{total} dispersion defined in SAPT, let alone be larger.
The D3 corrections for BLYP-D3 presented in \tabref{tab-disp} are therefore unphysical.
The spuriously large dispersion contribution is only somewhat reduced for B3LYP-D3.

\begin{figure}
\includegraphics[width=1.0\textwidth]{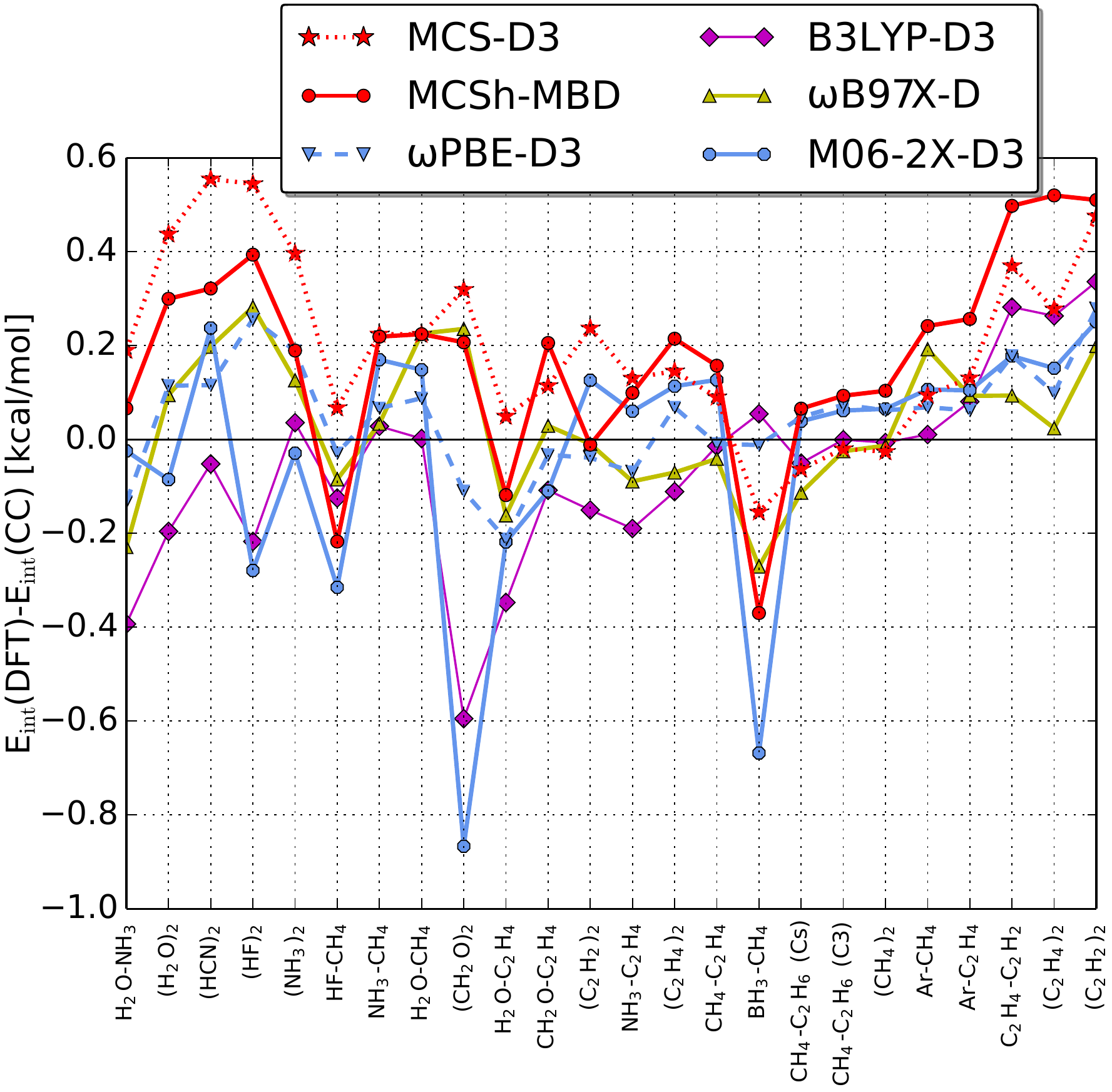}
\caption{Detailed results for the A24 database. The deviations are with
 respect to the nonrelativistic interaction energies at the CCSD(T)/CBS level plus
 CCSDT(Q) corrections.\cite{rezac2013describing} The energies for the functionals other
than MCS are taken from \citet{li2014quantum}} \label{fig-a24}
\end{figure}

\begin{table}
\caption{Statistical errors for the S22 and A24
 databases.\cite{jurecka2006benchmark,rezac2013describing}
The units are kcal/mol.} \label{tab-s22a24}
 \begin{tabular}{lllll}
\hline\hline
               & MAPD & RMSD & MAD  & MSD   \\
\hline
\bf{S22}       &      &      &      &       \\
MCS-D3         & 7.05 & 0.54 & 0.42 & 0.08 \\
MCS-MBD        & 6.05 & 0.47 & 0.34 & 0.07 \\
MCSh-D3        & 5.44 & 0.44 & 0.34 & 0.15 \\
MCSh-MBD       & 5.94 & 0.31 & 0.25 & 0.03 \\
$\omega$PBE-D3 & 6.65 & 0.36 & 0.27 & -0.11 \\
M06-2X         & 7.38 & 0.53 & 0.38 & 0.20 \\
M06-2X-D3      & 6.39 & 0.47 & 0.34 & -0.12 \\
B3LYP          & 86.4 & 4.91 & 3.76 & 3.76 \\
B3LYP-D3       & 6.68 & 0.48 & 0.39 & -0.20 \\
BLYP-D3        & 5.41 & 0.33 & 0.24 & -0.20 \\
$\omega$B97X-D & 7.37 & 0.32 & 0.23 & -0.18 \\
\hline
\bf{A24}       &       &      &      &        \\
MCS-D3         & 16.38 & 0.27 & 0.22 & 0.20 \\
MCS-MBD        & 23.67 & 0.30 & 0.27 & 0.24 \\
MCSh-D3        & 16.45 & 0.27 & 0.21 & 0.21 \\
MCSh-MBD       & 22.82 & 0.27 & 0.23 & 0.17 \\
$\omega$PBE-D3 & 8.06  & 0.12 & 0.10 & 0.05 \\
M06-2X         & 20.51 & 0.29 & 0.23 & 0.04 \\
M06-2X-D3      & 14.47 & 0.27 & 0.19 & -0.03 \\
B3LYP          & 99.6 & 1.10 & 1.00 & 1.00 \\
B3LYP-D3       & 9.58  & 0.21 & 0.15 & -0.06 \\
$\omega$B97X-D & 10.05 & 0.15 & 0.12 & 0.03 \\
\hline\hline
 \end{tabular}
\end{table}

\begin{table}
\caption{Comparison of the D3 dispersion correction and SAPT dispersion plus exchange-dispersion
for selected complexes from the S22 database. The SAPT dispersion energies are taken from \citet{pernal2009dispersionless}
The units are kcal/mol.}\label{tab-disp}
\begin{tabular}{lrrrr}
\hline\hline
dimer                   & SAPT        &  B3LYP-D3        & MCS-D3             & BLYP-D3 \\
\hline
\ce{(CH4)_2}            &-1.06        &-0.92             &-0.79                & -1.18   \\
\ce{(C_2H_4)_2}         &-2.58        &-2.12             &-1.52                &  -2.90  \\
uracil dimer stack      &-11.08       &-9.16             &-6.87                & -11.52 \\
\ce{C6H6-H2O}           &-2.82        &-2.32             &-1.73                & -2.89  \\
\ce{C6H6-NH3}           &-2.86        &-2.36             &-1.76                & -2.91  \\
\hline\hline
\end{tabular}
\end{table}

\subsection{Water clusters}
Water clusters constitute a challenge for approximate DFT methods. Although water molecules are polar,
their clusters are bound not only by electrostatics and induction, but also largely by the dispersion
effects. More importantly, the clusters sample interactions not represented in
the standard test databases: interactions with distant neighbors and multiple hydrogen bonds formed by a single water molecule.

\begin{figure}
 \includegraphics[width=1.0\textwidth]{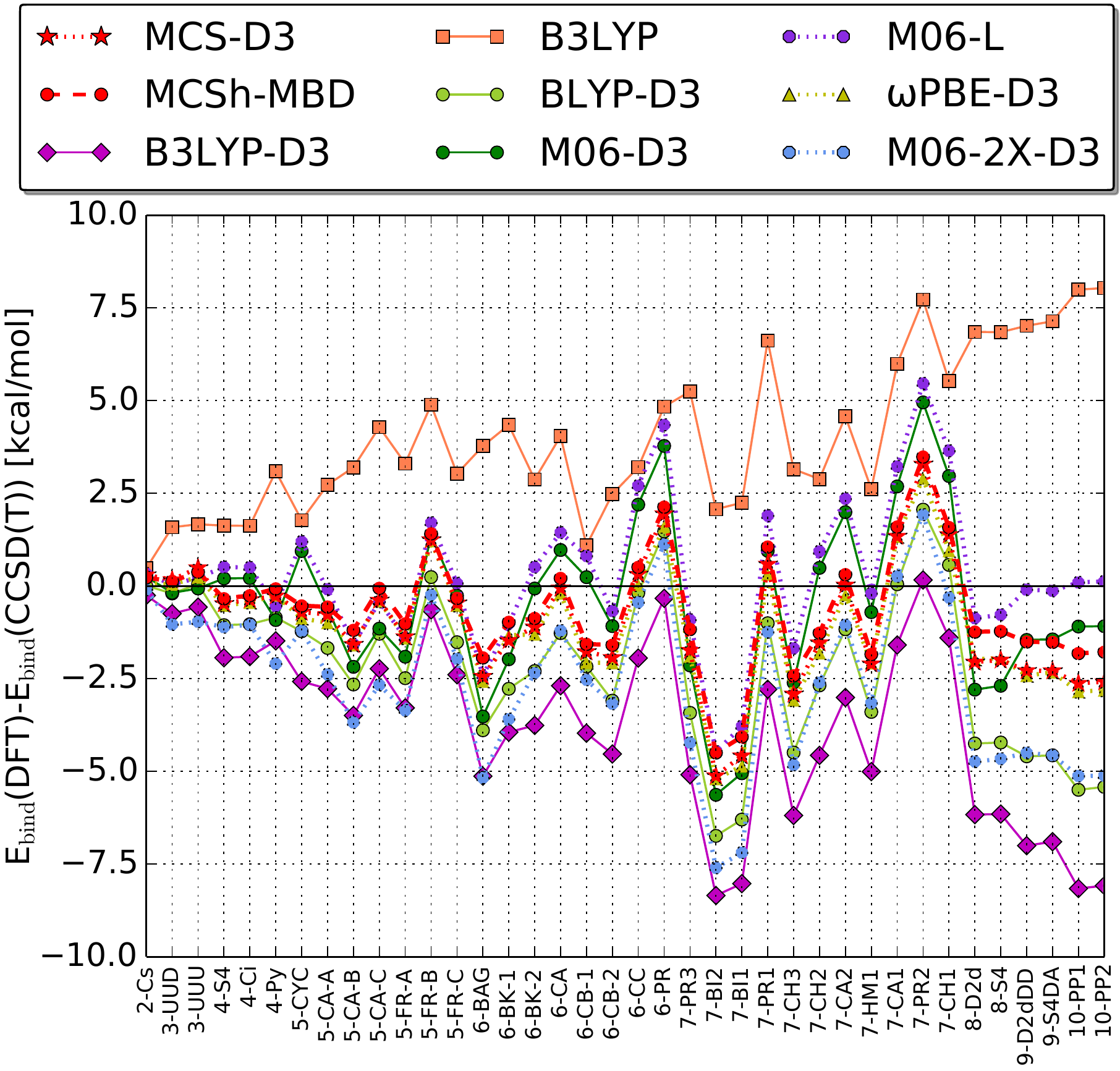}
 \caption{Differences between the CCSD(T)/CBS and DFT binding energies for \ce{(H2O)_n} with
 $n = 2, \ldots, 10$. The coordinates, reference energies, and labels of the water clusters
 are taken from \citet{temeleso2011benchmark}} \label{fig-shields}
\end{figure}

Water clusters exemplify the advantage of our approach over the dispersion-corrected functionals
based on massive error cancellation. \figref{fig-shields} shows the system-size dependence of the errors
of various methods. The MCS functionals show no
systematic underbinding or overbinding. This is in contrast to the functionals based on the B88 exchange: B3LYP
systematically underbinds, while both B3LYP-D3 and BLYP-D3 systematically overbind due to the overcorrection of
the B88 exchange by the D3 term. This error cancellation had no adverse effects in the previous test cases.

\tabref{table-shields} illustrates that all four MCS functionals yield exceptionally
 small relative and absolute errors for \ce{(H2O)_n} with
 $n = 2, \ldots, 10$. While the choice of the dispersion correction does not
influence the average errors, the choice of the exchange functional is more important.
The hybrid MCS functionals perform significantly better than the pure counterparts. Of all the tested functionals, 
MCSh-MBD offers the best performance for the water clusters of \figref{fig-shields}.

\begin{table}
  \caption{Statistical errors of DFT methods for \ce{(H2O)_n} with
 $n = 2, \ldots, 10$. The units are kcal/mol.} \label{table-shields}
  \begin{tabular}{lllll}
    \hline\hline
    Functional & MAD & MAPD & MSD & RMSD \\
    \hline
    MCS-D3      & 1.53 & 3.17 & -0.94 & 1.92 \\
    MCS-MBD     & 1.52 & 3.12 & -1.02 & 1.91 \\
    MCSh-D3     & {\bf 1.25} & 2.92 &  0.87 & 1.69 \\
    MCSh-MBD    & 1.28 & {\bf 2.65} & -0.59 & {\bf 1.64} \\
    M06-D3      & 1.71 & 3.58 & -0.43 & 2.22 \\
    M06-2X-D3   & 2.75 & 5.79 & -2.58 & 3.35 \\
    B3LYP       & 4.01 & 8.53 &  4.01 & 4.52 \\
    B3LYP-D3    & 3.66 & 7.39 & -3.65 & 4.39 \\
    BLYP-D3     & 2.42 & 4.68 & -2.18 & 3.04 \\
    {\hjswpbed} & 1.55 & 3.06 & -1.09 & 1.97 \\
    M06-L       & 1.40 & 3.13 &  {\bf 0.29} & 1.98 \\
    \hline\hline
  \end{tabular}
\end{table}

\begin{figure}
 \includegraphics[width=1.0\textwidth]{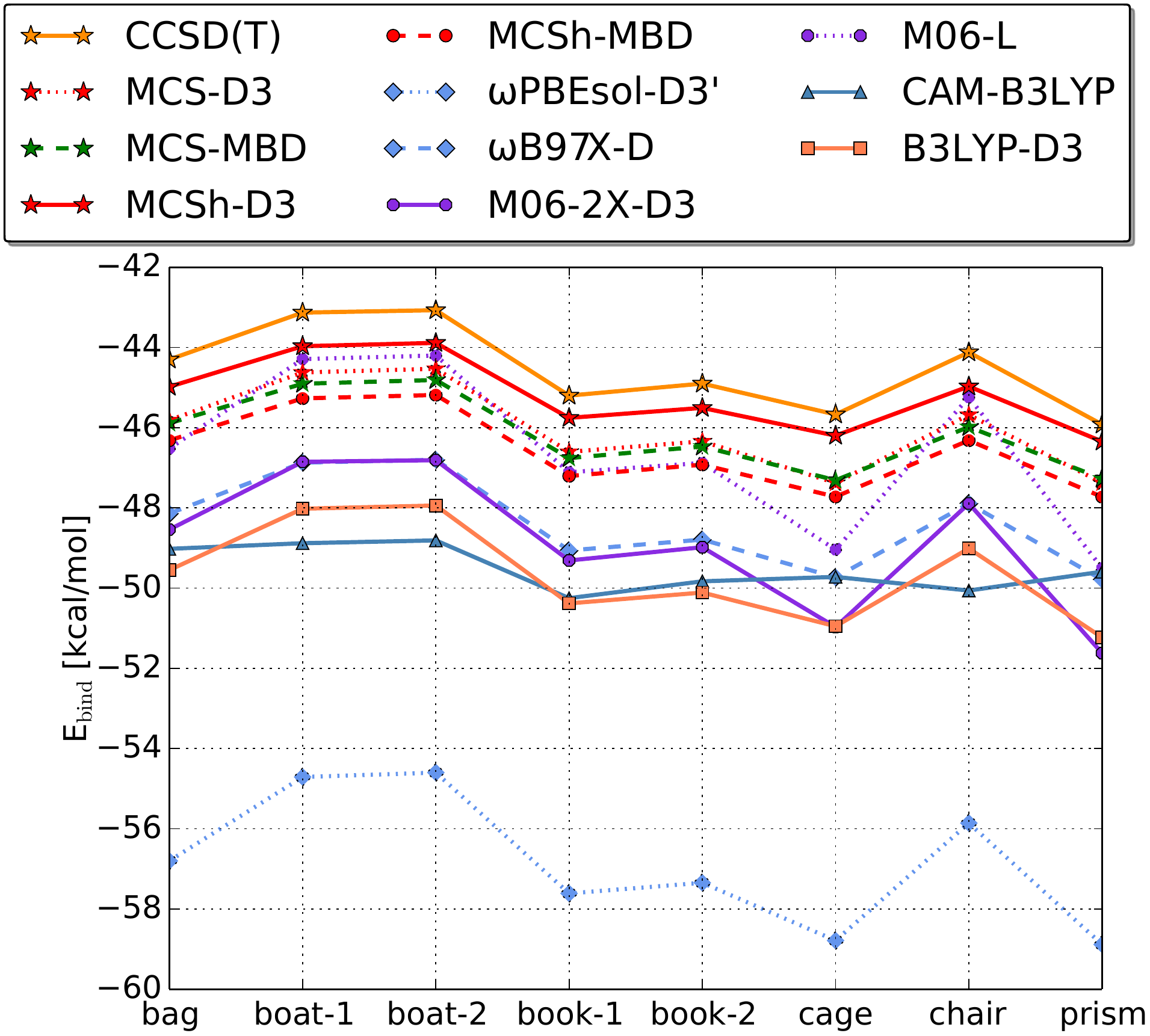}
 \caption{Binding energies of water hexamers. The coordinates and reference CCSD(T)/CBS energies 
are taken from \citet{bates2009ccsd}. The energies for the functionals other than MCS are taken
 from \citet{leverentz2013assessing}} \label{fig-6mers}
\end{figure}

\figsref{fig-6mers}{fig-16mers} focus on the performance of approximate methods for
water hexamers and 16-mers. All the MCS functionals yield small absolute errors, excellent ordering
 of the hexamers, and well reproduced (although not perfectly)
tiny energy differences between the 16-mers. The effect of changing the dispersion correction
is negligible when the exchange is pure ($\alpha=0$). However, there is an appreciable difference
between MCSh-D3 and MCSh-MBD for the water 16-mers. While the binding energy for MCSh-MBD agrees
almost perfectly with the reference values, MCSh-D3 underbinds by about \SI{4}{kcal\per mol}.

The excellent performance of the MCS functionals for the 16-mers is encouraging because
these systems exhibit features that are expected in even larger clusters. First, among the systems
considered in this study, only the 16-mers contain  water molecules participating in four hydrogen bonds. 
Moreover, the energetics of the 16-mers include significant many-body effects, which are large compared
to the energy differences between the isomers. Indeed, \citet{wang2013benchmark} have estimated that the 5-body
and higher effects in the 4444-a 16-mer to contribute \SI{-2.3}{kcal\per mol}
 to the binding energy at the MP2 level. This is a highly probable estimate since in our computations the MP2
 method is shown to approach extremely close the CCSD(T)/CBS limits for all 16-mers (see \tabref{tab-16mers}).

Many-body effects in the 16-mers are dominated by induction terms, as shown by several studies
 on trimers of polar molecules.\cite{szczesniak1992initio,szczesniak1991initio,chalasinski1991initio}
This explains why the performance of MP2 is excellent for the 16-mers despite inability of MP2 to recover
the third-order triple-dipole dispersion terms. The induction nature of many-body effects
justifies the D3 atom-pairwise dispersion correction, which does not comprise any nonadditive
 three-body dispersion terms.\cite{grimme2010consistent} In fact, we have not observed any significant
improvement attributable to the MBD dispersion correction which is capable of recovering many-body
dispersion.

It should be emphasized that our reference binding energies of the 16-mers
are uniformly shifted with respect to those used by \citet{leverentz2013assessing}
This is because these authors employed the CCSD(T)/aug-cc-pVTZ energies of \citet{yoo2010high}
as their final reference values, whereas in our study these energies have been refined
in the extrapolation scheme defined in \eqnref{ccsdt-extrap}. The extrapolation has introduced
an upward shift of about \SI{6.5}{kcal/mol} relative to CCSD(T)/aug-cc-pVTZ. A recent quantum Monte
Carlo result of \citet{wang2013benchmark} for the 4444-a isomer (\SI{-165.1(8)}{kcal\per mol})
is in excellent agreement with our CCSD(T)/CBS extrapolation (\SI{-164.51}{kcal\per mol}).

\begin{table}
 \caption{Binding energies of water 16-mers. The CCSD(T) and RI-MP2 energies are extrapolated
according to \eqnref{ccsdt-extrap} and \eqnref{mp2-extrap}, respectively. The units are kcal/mol.}\label{tab-16mers}
 \begin{tabular}{lrrrrr}
\hline\hline
system  &CCSD(T) &RI-MP2  & MCS-D3 & MCSh-MBD \\
\hline
4444-a  &-164.51 &-163.91 &-166.54 & -164.55 \\
4444-b  &-163.97 &-163.46 &-166.37 & -164.24 \\
antiboat&-164.11 &-164.07 &-166.48 & -164.88 \\
boat-a  &-164.40 &-164.45 &-166.99 & -165.36 \\
boat-b  &-164.28 &-164.33 &-166.79 & -165.17 \\
\hline\hline
 \end{tabular}
\end{table}

\begin{figure}
 \includegraphics[width=1.0\textwidth]{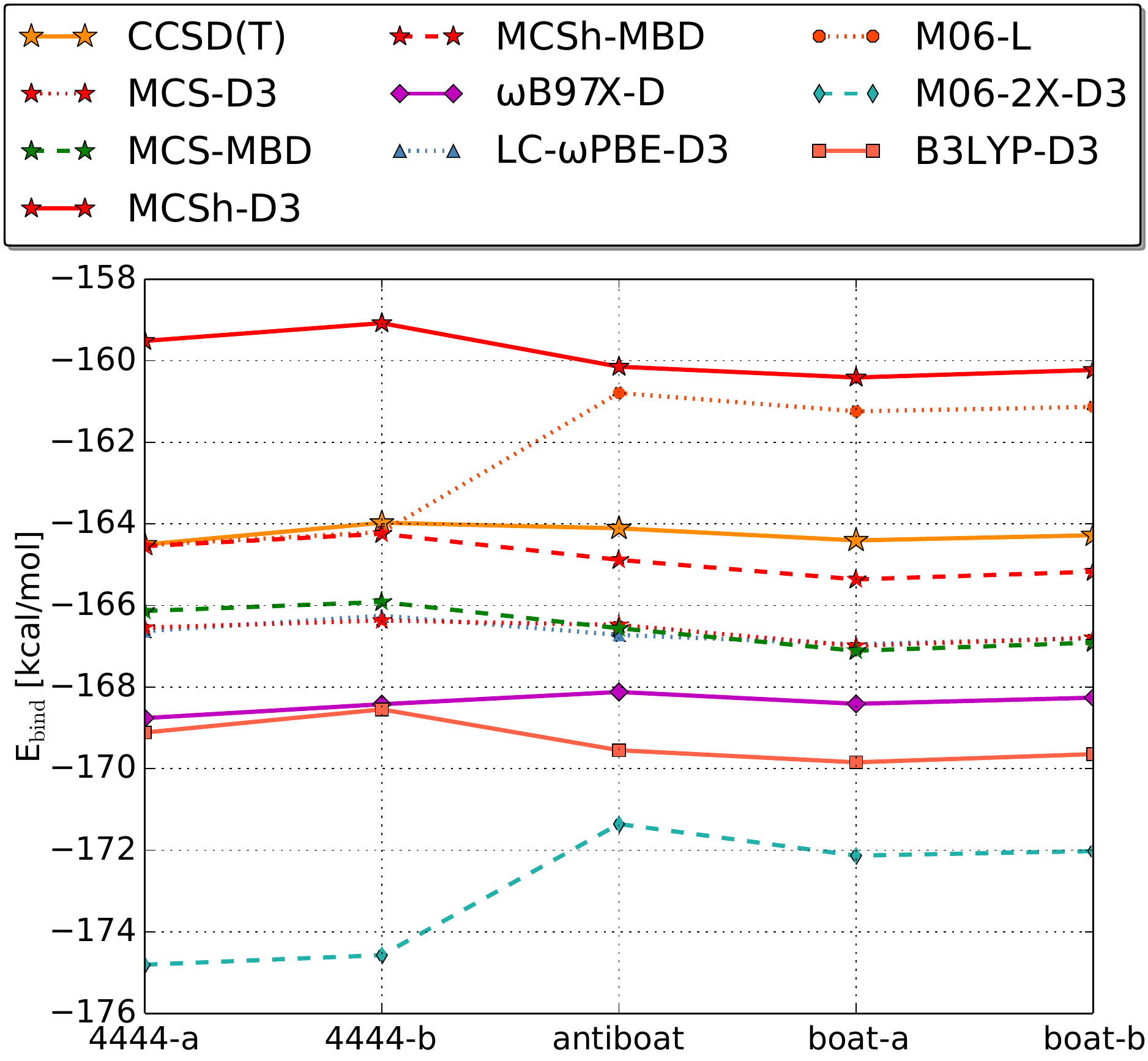}
 \caption{Binding energies of water 16-mers. The coordinates are taken from \citet{yoo2010high}. 
 The energies for the functionals other than MCS
 are taken from \citet{leverentz2013assessing}.} \label{fig-16mers}
\end{figure}

\subsection{Ionic hydrogen bonds}
Hydrogen-bonded systems composed of an ion interacting with a closed-shell molecule provide
a simple model of interactions ubiquitous in biochemistry. From the point of view of dispersion corrections,
charged dimers belong to the hardest cases: if a dispersion correction does not depend on the density,
it will not reflect any alterations of dispersion due to the density change from a neutral to an ion,
which is often dramatic. This is the case of the D3 model which has been parametrized within a set of neutral
dimers, and its input consists of atomic coordinates only.\cite{grimme2012performance}
However, because the total interaction is dominated by electrostatic and induction components,
this weakness may not be especially relevant, as dispersion itself is relatively small and thus its accuracy
not critical.

\begin{figure}
\includegraphics[width=1.0\textwidth]{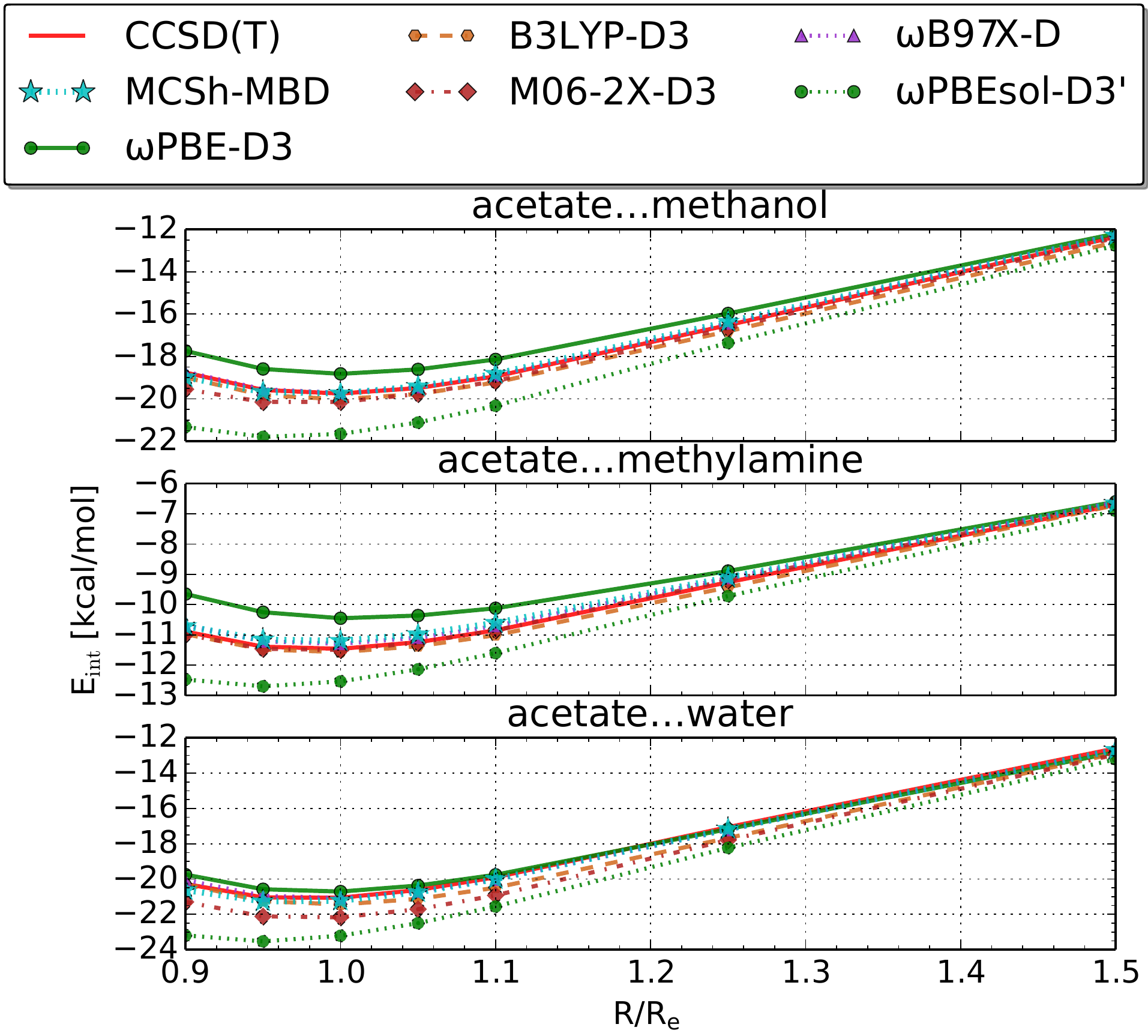}
\caption{Interaction energy curves for hydrogen-bonded dimers including the acetate anion.
The data for B3LYP-D3, M06-2X-D3, and B97X-D are taken from Ref.~\citenum{li2014quantum}.} \label{fig-acetate}
\end{figure}

\figsref{fig-acetate}{fig-imidazolium} show the performance of MCSh-MBD and compare this functional with
the results of popular DFT methods. The differences between MCSh-MBD, B3LYP-D3, M06-2X-D3,
and $\omega$B97X-D are small, and all of the curves are close to the reference ones. 
The $\omega$PBE-D3 functional is consistently worse than any of the MCS functionals (see \tabref{tab-ionic})
despite its good performance for water clusters.

\tabref{tab-ionic} shows that switching from D3 to MBD changes little when applied with the pure MCS functionals.
However, the choice the dispersion correction appears more important for the hybrid variants,
and the MBD model works better in this case. This observation is consistent with our findings for the
hydrogen-bonded dimers of the S22 database and for water clusters.

\begin{table}
 \caption{Interaction energies of ionic hydrogen-bonded dimers at the equilibrium
 distances. The units are kcal/mol.}\label{tab-ionic}
 \begin{tabular}{lllllll}
\hline\hline
 dimer           & CCSD(T)    & MCS-D3 & MCS-MBD & MCSh-D3 & MCSh-MBD & {\hjswpbed} \\
\hline
 \multicolumn{7}{l}{\bf acetate\ldots X}  \\
 methanol        & -19.75     & -19.46 & -19.57	 & -19.28  & -19.74   & -18.82 \\
 water           & -21.06     & -20.97 & -21.15  & -20.73  & -21.22   & -20.71 \\
 methylamine     & -11.46     & -10.96 & -11.01  & -10.85  & -11.20   & -10.45 \\
  \multicolumn{7}{l}{\bf imidazolium\ldots X} \\
 formaldehyde    &     -16.41 & -15.84 & -15.86  & -15.94  & -16.14   & -15.75  \\
 methylamine     &     -25.98 & -26.58 & -26.62  & -26.69  & -27.04   & -26.86  \\
 water           &     -16.49 & -16.25 & -16.30  & -16.30  & -16.54   & -15.89  \\
\hline\hline
 \end{tabular}
\end{table}

\begin{figure}
\includegraphics[width=1.0\textwidth]{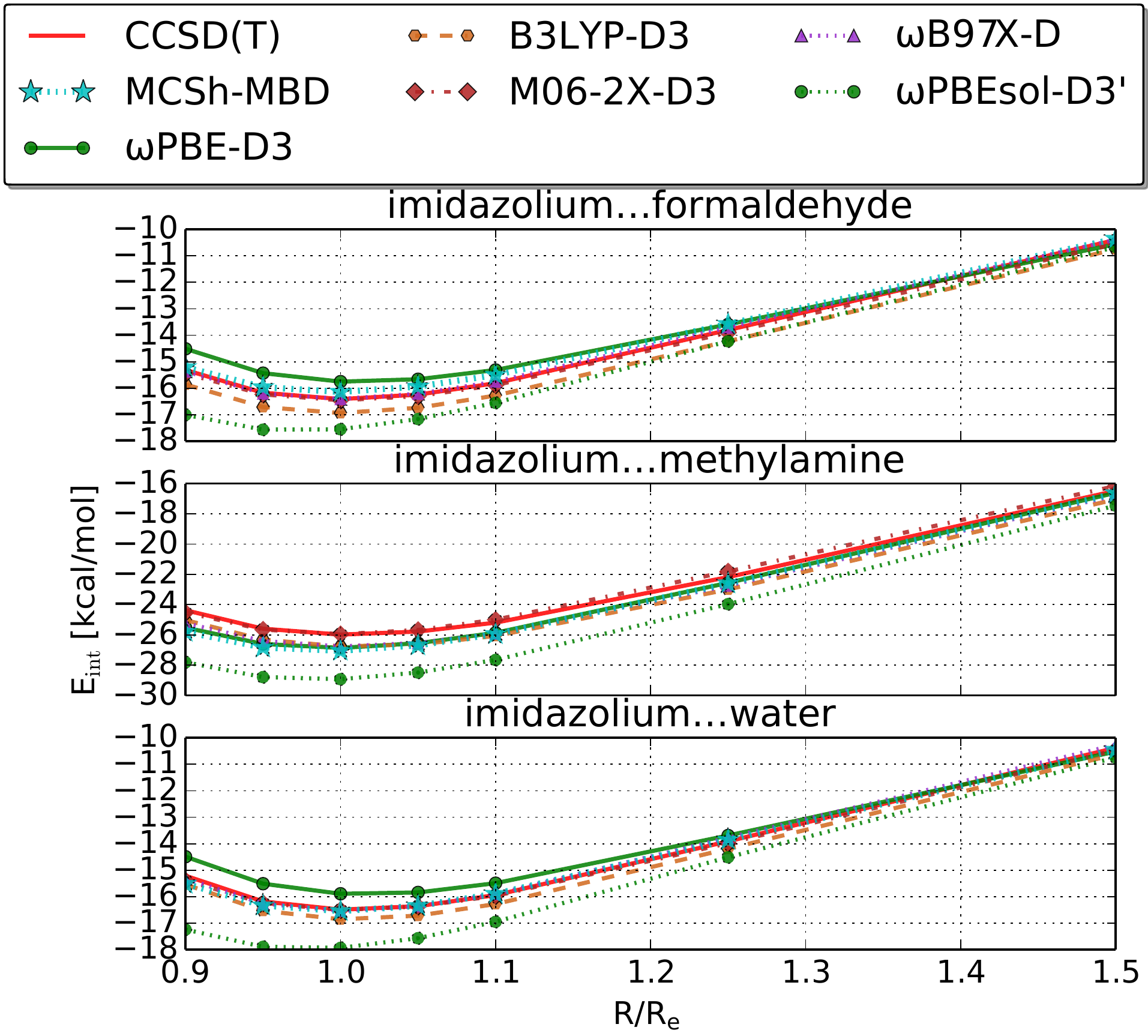}
\caption{Interaction energy curves for hydrogen-bonded dimers including the imidazolium cation.
The data for B3LYP-D3, M06-2X-D3, and B97X-D are taken from Ref.~\citenum{li2014quantum}.} \label{fig-imidazolium}
\end{figure}

\subsection{Isodesmic reaction of n-alkanes}
The systematic errors of DFT approximations in predicting alkane thermochemistry were
discussed by \citet{wodrich2006systematic}, \citet{song2010calculations},
and \citet{grimme2010alkane}. They observed that there is a substantial error
in reaction energies of isodesmic ethane fragmentation reactions of alkanes,
which accumulates as the chain length grows,
\begin{equation}
\ce{CH3(CH2)_{$m$}CH3 + $m$CH4 -> $(m+1)$ C2H6}. \label{isodesmic-reaction}
\end{equation}

The performance of approximate functionals for these reactions
is connected to the quality of the description of noncovalent interactions. \citet{johnson2012density} found that
the error in the reactions of \eqnref{isodesmic-reaction} has its origin in the region of space between 1,3 methylene groups where
the reduced density gradient changes upon fragmentation of an alkane to ethane. This change is a signature of
noncovalent bonds.\cite{johnson2010revealing} 

\begin{figure}
 \includegraphics[width=1.0\textwidth]{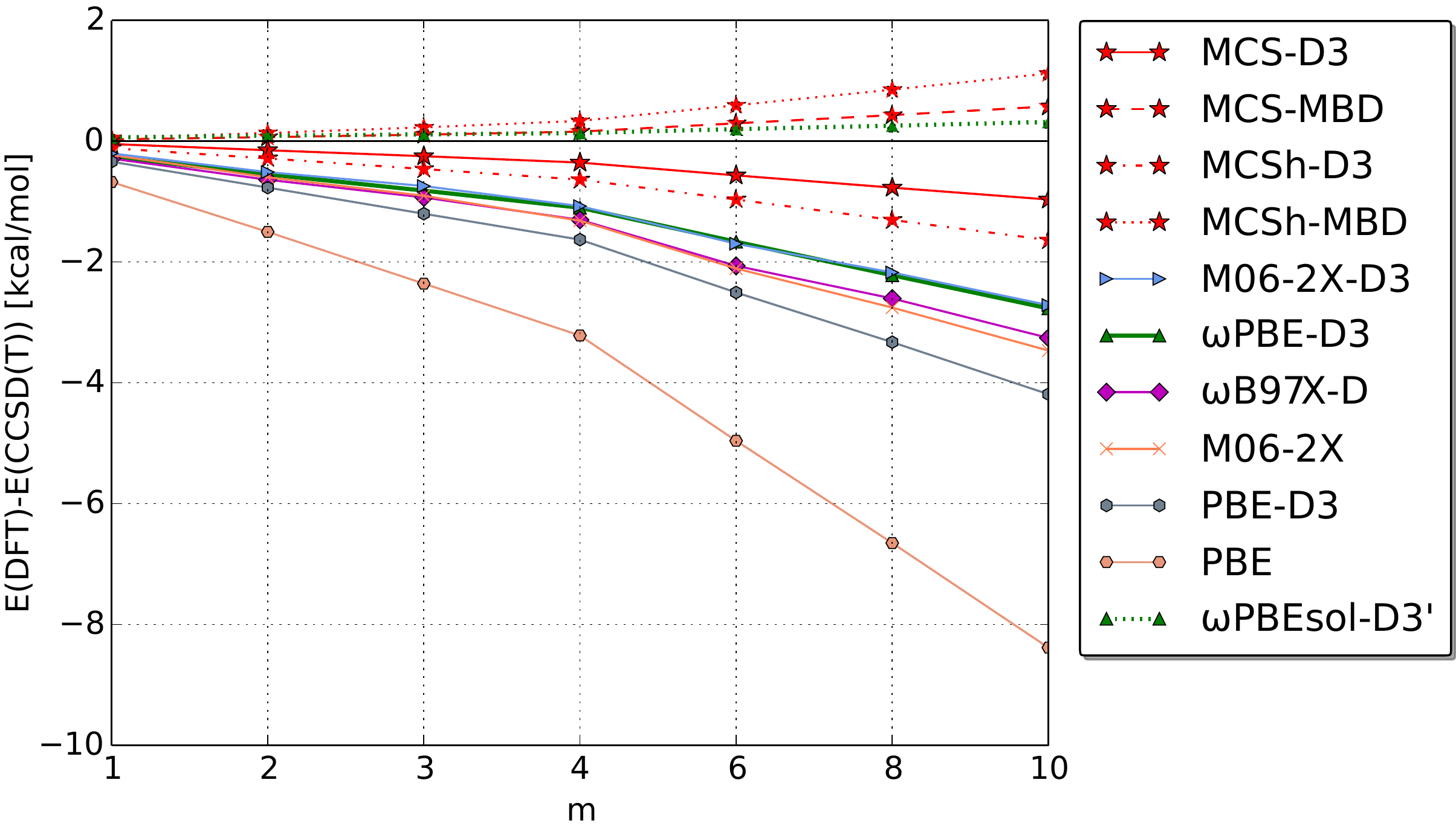}
 \caption{Errors in the energy of the isodesmic reaction.
 The reference energies and coordinates are taken from \citet{grimme2010alkane}.} \label{fig-isodesmic}
\end{figure}

Previous studies identified the features that a functional should possess
to alleviate this problem: \begin{inparaenum}[(i)]
\item range-separation of the exchange functional,\cite{song2010calculations} \item restoration of the exact gradient
expansion of the exchange\cite{csonka2008improved} (as in PBEsol), \item a dispersion
correction.\cite{song2010calculations,grimme2010alkane} \end{inparaenum}
The MCS functionals as well as $\omega$PBEsol-D3' (discussed in the next section) include all of the above features.
As shown in \figref{fig-isodesmic}, these methods are by far the best performers for reactions in question.

\subsection{Merit of the MCS correlation}
The question remains as to whether the correlation functional of our approach is indeed
crucial to the quality of the above presented results. One might argue that
this accuracy is primarily determined by the exchange and dispersion parts, and only 
weakly dependent on the semilocal correlation. To verify this hypothesis, we have
composed a functional which differs from MCS-D3 only by the PBEsol correlation (denoted
as $\omega$PBEsol-D3'), that is, both MCS-D3 and $\omega$PBEsol-D3' share the same
$\omega$PBEsol exchange with $\omega=0.3$ and the same D3 correction. \figref{fig-6mers}
shows that keeping the PBEsol correlation leads to ca.~$25\%$
overbinding in the case of water hexamers. A similar overbinding occurs for the 16-mers
(e.g. $E_\mathrm{bind}=\SI{-206.1}{kcal\per mol}$ for the isomer 4444-a). Furthermore, $\omega$PBEsol-D3'
overestimates the interaction energies for every ionic hydrogen-bonded dimer presented in \figsref{fig-acetate}{fig-imidazolium}.
Evidently, the role of the MCS correlation is essential
in these examples, and its replacement by the standard PBEsol correlation leads to serious
overestimation of interaction energies.

Nonetheless, it is of note that there exist cases where the choice of a semilocal correlation
part matters less. For alkane fragmentation reactions, \figref{fig-isodesmic}, $\omega$PBEsol-D3'
performs even better than the MCS functionals, which suggests the dominant role of
the $\omega$PBEsol exchange in this case.

\section{Summary and conclusions}
We have proposed a new DFT exchange-correlation functional that is
specifically optimized  for noncovalent interactions. It is composed
of well-defined and physically meaningful components, with minimum of empiricism
and reduced opportunity for error cancellation. It is built of the meta-GGA correlation functional
developed by \citet{modrzejewski2012first} and the range-separated
PBEsol exchange. The exchange and correlation contain a slight amount of empiricism
in a form of parameters defining the scope of various approximations:
a single parameter which governs damping of the semilocal correlation hole at large $r_{12}$,
a range-separation parameter controlling the onset of the long-range HF exchange,
 and---in the case of the hybrid exchange---a fraction of the short-range HF exchange.

The novel piece of our functional, the correlation functional, is designed with the constraint satisfaction technique,
but with the aid of its single empirical parameter it may be finely adjusted
to any accurate variant of a dispersion correction without compromising
any formal or physical constraints that it satisfies.

We have calibrated two long-range dispersion corrections to work with the remaining part of the functional:
 D3\cite{grimme2010consistent} and MBD.\cite{ambrosetti2014long}
Taking into account the two possible variants of exchange and the two variants of dispersion,
there is a set of four MCS functionals which are tested in this study.

The test set is composed of popular databases of small noncovalent dimers,
but includes also the more demanding cases of water clusters,
hydrogen-bonded interactions in ion-neutral pairs, and the thermochemistry of
isodesmic reactions of n-alkanes. For the classic S22 database, the MCS functionals
 perform on a par or better than the leading functionals in the field of
 noncovalent interactions: B3LYP-D3, M06-2X-D3, and $\omega$B97X-D.
More importantly, the MCS functionals perform markedly better than these functionals for 
large water clusters for which they successfully predict the binding
energies from the newly refined CCSD(T)/CBS benchmarks. The good performance for
hydrogen bonding extends to ionic hydrogen bonds. Finally, all four MCS functionals
 display excellent performance in predicting the energetics of isodesmic
reactions of n-alkanes in direct consequence of the good description of the
intramolecular interactions between methylene groups. We find that the PBEsol exchange
combined with range separation and a dispersion correction essentially solves
the known problems of DFT with isodesmic reactions of alkanes.

In view of the presented results, all four MCS functionals could be recommended
for the description of noncovalent systems. The best performer in any case except
few-atom dispersion-bound systems is MCSh-MBD.

\section{Acknowledgments}
This work was supported by the National Science Foundation (Grant No.~CHE-1152474) and by 
the Polish Ministry of Science and Higher Education (Grant No.~N204~248440). 
M.M. and G.C. gratefully acknowledge additional financial support from the Foundation for Polish Science.
Special thanks to Aleksandra Tucholska for creating the TOC graphic for this paper.

\begin{figure}[p]
\includegraphics[width=0.7\textwidth]{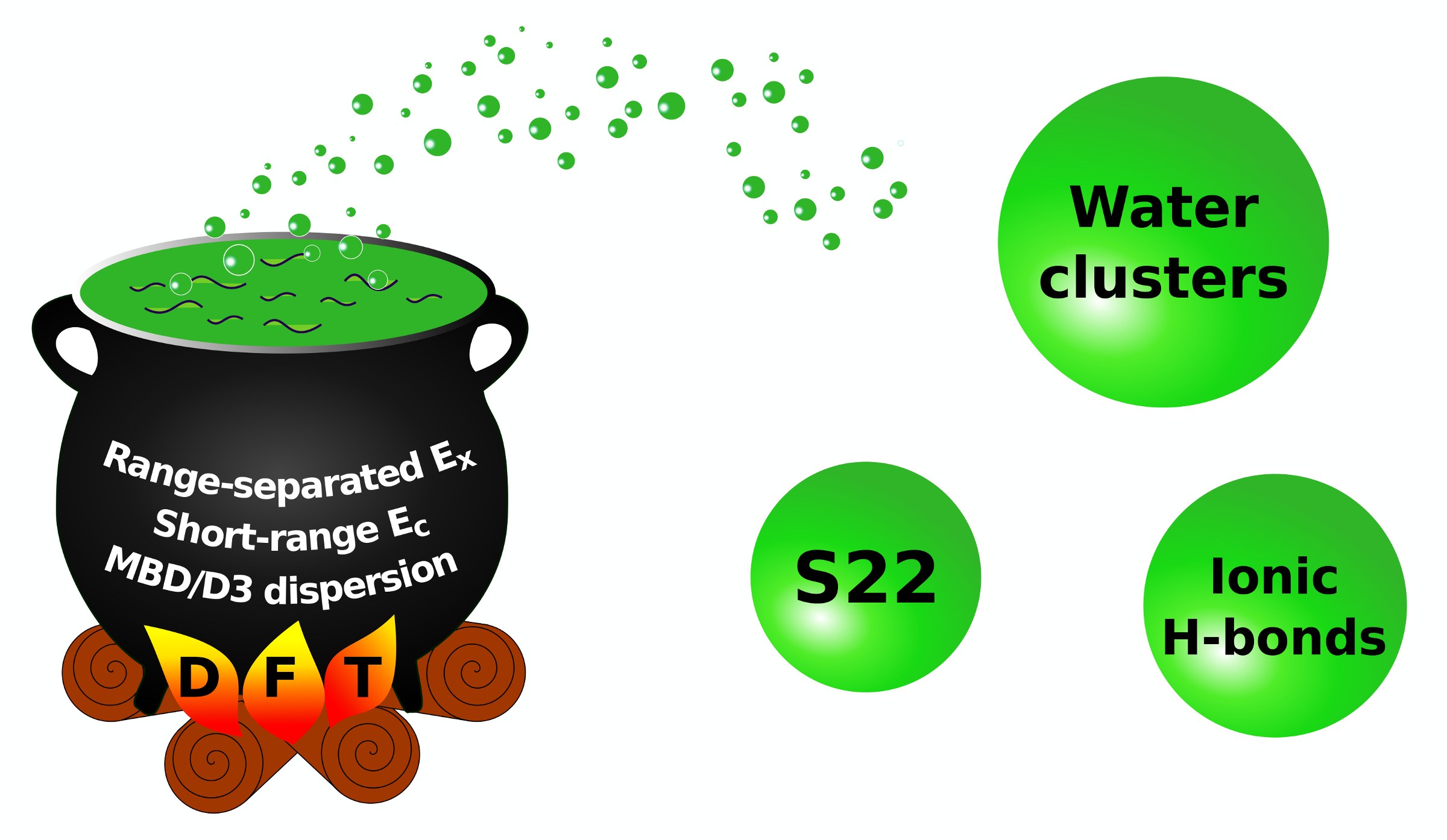}
\caption{For table of contents only}
\end{figure}

\bibliography{biblio}
\end{document}